\documentstyle[sprocl]{article}

\def\be{\begin{equation}}
\def\ee{\end{equation}}
\def\bea{\begin{eqnarray}}
\def\eea{\end{eqnarray}}

\def\d{\partial}
\def\bmat{      \left |  \begin{array}{cc} }
\def\emat{ \end{array} \right |    }

\begin{document}

\title{
POMERON and  AdS/CFT CORRESPONDENCE FOR QCD}

\author{Chung-I Tan}

\address{Physics Department, Brown University,
       Providence RI 02912, USA, E-mail: tan@het.brown.edu }


\maketitle\abstracts{ The Maldacena  conjecture that QCD is
holographically dual to gravity in extra dimensions is briefly reviewed.
On the basis of this duality conjecture, the complete glueball
spectrum is computed which bares a striking
resemblance to the known $QCD_4$ spectrum as determined by lattice
simulations.  In particular, a strong 
		coupling expansion for the Pomeron intercept is obtained.}

\bibliographystyle{unsrt}    


\section{Introduction}

It has been a long held belief that QCD in a non-perturbative setting
can be described by a string theory. 
This  thirty-year search for the QCD strings has recently led to a
remarkable conjecture: QCD is holographically dual to gravity in extra
dimensions. To be more precise, the Maldacena conjecture and
its further extensions state that Yang-Mills theory is exactly
dual to a critical string theory in a non-trivial gravitational
background~\cite{maldacena}. In such a framework, Pomeron  should
emerge as a closed string excitation~\cite{tan}.

In this talk, we provide a brief review  for the
Maldecena duality  conjecture, and summarize results for the glueball
spectrum and the Pomeron intercept in the strong coupling limit. 
The present discussion is  heuristic, disregarding important details
in order to provide a picture for the relevant physics.

Let us begin by first recalling that in the early days of string
theory, (or the ``dual resonance model'' to use the nomenclature that
predates both string theory and QCD), one observed that it was reasonable  to
represent the hadronic spectrum beginning with zero width ``resonances'' on
exactly linear Regge trajectories. With the advent of QCD this
approach was reformulated as the $1/N$ expansion at fixed 't Hooft
coupling, $g^2_{YM} N$. States with vacuum quantum numbers could be assigned to
closed-strings, including a massive
$2^{++}$ tensor glueball on the leading Pomeron trajectory, 
$
\alpha_P(t) = \alpha_{P}(0) + {\alpha'_P} \> t. 
$
Soon a three-fold crisis appeared: 
\begin{itemize}
\item{zero-mass states,}
\item{ extra dimensions,}
\item{supersymmetries.}
\end{itemize}
  A careful study of negative norm
states (i.e ghosts), tachyon cancellation and the consistency of the
perturbative expansion at the one loop level led to supersymmetric
string theories in 10 space-time dimensions. At the one-loop level
unitarity requires that pair creation of two open strings, each
contains ``zero-mass" spin-1 states, is dual to a vacuum exchange
 with  an intercept
$\alpha_P(0)=2$. This leads to a massless
$2^{++}$ state, the {\bf graviton}. In fact the low energy theory was
clearly not QCD but rather {\bf supergravity in 10 dimensions}! 

What is the mechanism which allows our 4-d space/time and yet
is able to generate a non-zero mass gap for tensor glueballs? How can
one ``lower" the Pomeron intercept so that
$\alpha_P(0)$  takes on its phenomenological value of
$1.1\sim 1.2$? The key ingredient turns out to be {\bf duality},
which allows a dual description of QCD involving  extra dimensions
and a nontrivial background metric which breaks supersymmetries.

On the basis of Maldacena's duality conjecture, a rich glueball
spectrum can be  computed at strong coupling.  This recent
development has led to the rebirth of active QCD string studies.

\section{ AdS/CFT duality for QCD}

Indeed, there is a trivial kinematic advantage in having (at least)
one extra dimension.  A zero-mass graviton in 5-d has 5 (rather than
2) on-shell states, [the  little group for a lightlike vector $p =
(p_\mu,p_5) = (p_5,0,0,0,p_5)$ is  $SO(3)$]. Thus a single 5-d
graviton can mimic the five spin-components for a 4-d massive
$2^{++}$ glueball on the leading trajectory, if it propagates through
a non-trivial ``medium". It is then possible that  a graviton in 5-d,
when seen from a 4-d perspective, becomes massive, thus leading to a
Pomeron with an intercept less than 2.

 This simple observation is at
the heart of the modern approach.  Maldacena's
conjecture begins with a full 10-d critical superstring with the
backgrond ``medium'' provided dynamically as a solution to
appropriate set of supergravity Einstein's equation.  The
self-consistency of this approach presumably is critical to the still
poorly understood mathematical framework required to prove Maldecena's
duality conjecture.

\subsection{Mass Generation and Extra Dimensions}
Before mentioning some technical details, let
us demonstrate how the incorporation of extra-dimensions with a
nontrivial background provides a natural mechanism for
mass-generation. Consider a  situation where the addition of a
fifth-dimension, r, leads to a metric, $ ds^2 =  u(r)^2 \sum_{i=1,2,3,4}
d x_i^2 + w(r)^{-2} d r^2 $, where $w(r)$ takes on an appropriate
non-trivial form to be specified later. The  wave equation for a
minimally coupled massless  scalar field  is:
$
\big\{ \d_{\mu}^2 +\sigma(r)^{-1} \d_r \tau(r) \d_r \big\}\phi(x,
r)=0,
$
where $\tau(r) = u^4 w$ and $\sigma (r) = u^2 w^{-1}$.  If one
attempts  to send a 4-d plane-wave with 4-momentum
$p_\mu$, i.e., $\phi(x,r)=e^{ip\cdot x} \phi(r)$, one finds that 
$\phi(r)$ satisfies an ordinary diferential equation
\be
 -  \d_r \tau(r) \d_r \phi(r)=m^2 \sigma (r)  \phi(r),
\ee
 with $-p^2=m^2$. This equation is of the standard Sturm-Liouville
form, and the allowed $m^2$ values  can be found by treating this as
an eigenvalue problem.  If 
$\tau(r)$ and
$\sigma(r)$ could be chosen so that no  massless mode exists, one
would have massive propagation from a 4-d perspective.

\subsection{AdS/BH  Background for QCD}

One example of this approach begins with type-IIB string theory
leading to $QCD_3$ strings.  We know that there are many extended
solitonic objects beyond the perturbative string expansion.  In
particular in IIB, there are $3+1$ dimensional objects called Dirichlet
3-branes, (D3-branes), and the low-energy dynamics of a set of N
parallel D3-branes is described by  open strings with end-points
restricted on these branes. Indeed the effective theory (or
Born-Infeld action) reduces exactly to $SU(N)$ Yang-Mills theory at
weak coupling.  (To be exact, in the present context, the full target
gauge theory is ${\cal{N}} = 4$ SUSY YM in 4-d.)
Note that one set of zero-mass states, the spin-1 {\bf gauge bosons},
is no longer an embarassment as they represent the weakly coupled
gluonic modes. Because of branes, YM theories live in dimensions less
than 10.

The Maldacena conjecture states that both the open string/Yang-Mills
and the closed string/gravity descriptions are {\bf simultaneously}
true or equivalent in the near horizon limit, where the above metric
is $AdS^5 \times S^5$. In this limit, the theory does not contain
gravity, i.e., massless graviton {\bf decouples.}

However this background is so symmetric that the
resultant target Yang-Mills theory is conformal with
${\cal N} = 4$ supersymmetries.  Witten~\cite{maldacena} has suggested a
further modification to remove these unwanted symmetries. One
imagines raising the ``temperature'' by compactifying one 
spatial co-ordinate on a circle, $S^1$, parallel to the brane, with
anti-periodic fermionic boundary conditions to break SUSY.  In the
gravity language, the resultant background metric collapses into an
$AdS^5$/black-hole, while on the gauge side the 4-d theory deconfines
and lifts both the fermion masses and through quantum corrections the
scalar masses to the ``cut-off'' scale provided by this temperature. 
At energies well below the cut-off in the weak coupling region, it is
conjectured that one is left with a dimensionally reduced pure 3-d
Yang-Mills theory or quarkless $QCD_3$. 

To arrive at $QCD_4$ as the
target theory, one begins with the eleven dimensional M theory on
$AdS^7 \times S^4$ or 10-d type-IIA string theory after compactifying
the ``eleventh'' dimension (on a very small circle of radius $R_0$).
Again following the suggestion by Witten, the ``temperature" is
raised with a second compact radius $R_1$ in a direction $\tau$
parallel to the type-IIA D4-branes.  On the ``thermal'' circle, the
fermionic modes have anti-periodic boundary conditions {\bf breaking
conformal and all SUSY symmetries}.  Therefore, if all goes as
conjectured, in the scaling limit $g^2 N = g_s N \beta/R_1
\rightarrow 0$ there should be a fixed point mapping type-IIA string
theory in a background $AdS^7$/black-hole metric,
\begin{equation}
 ds^2 = f(r)
d\tau^2 + r^2 \sum_{i=1,2,3,4,11} d x_i^2 + f(r)^{-1} d r^2 +
d\Omega_4^2 \; , 
\end{equation} 
$f=r^2-{r^{-4}}$, into pure $SU(N)$ Yang-Mills  or quarkless $QCD_4$. 
Therefore, the threefold crisis mantioned earlier {\bf has now been
resolved.}

\section{Wave Equations and Glueball Spectra}

To compute the glueball excitations for $QCD_4$, in the
extreme strong coupling limit, one simply needs to find the spectrum
of harmonic fluctuations for the bosonic supergravity fields around
these AdS/black-hole backgrounds. The ``warp factor'' in the radial
``fifth'' coordinate forms a ``cavity'' so that all modes are
discrete and massive.~\cite{BMT1,BMT2}

The goal is to compute all strong coupling states that might survive
for $QCD_4$ in the scaling (weak coupling) limit.  We
therefore ignore throughout any Kaluza-Klein modes in compact
manifolds (compactified $S^1$ co-ordinates in AdS or the Sphere $S^4$), 
which are charge states in their own
superselection sector. 
To illustrate the approach consider
 metric fluctuations about the fixed
$AdS^7$ background, 
$G_{\mu\nu} = {\bar g}_{\mu\nu} + h_{\mu\nu}(x) ,$ where   
\be
h_{\mu\nu}(x) = H_{\mu \nu}(r) e^{i k_4 x_4} \; .
\ee
With all other fields set to zero, we look for plane wave solutions
with Minkowski time, $t = i x_4$ and discrete mass eigenvalues,
$m = -i k_4$.

 Because of
an accidental $SO(4)$ symmetry in $(x_1,x_2,x_3, x_{11})$ in strong
coupling, the spectrum is degenerate when classified by spin under
$SO(3)$. Here we exhibit explicitly the fluctuations leading to spin-2
tensor glueballs.  There are five independent perturbations,
$h_{ij}=q_{ij} r^2 T_4(r) e^{-mx_4}$, which form the spin-2
representations of $SO(3)$, where $i,j=1,2,3$ and $q_{ij}$ is an
arbitrary constant traceless-symmetric $3\times 3$ matrix. $T_4(r)$
satisfies the free wave equation,
\begin{equation}
r(r^6 - 1) T_4''(r) + (7 r^6 - 1) T_4'(r) + (m^2r^3) T_4(r)= 0.
\end{equation}
  This equation can be expressed covariantly,
as if one is dealing with  a minimally coupled
massless scalar field. In particular, the mass for the lowest tensor
glueball can be obtained:
$ m_T \simeq [9.86 + 0( \frac{1}{g^2 N} )] \;
\beta^{-1} \; ,
$
where $\beta=2\pi R_1$, with $R_1$ being  the thermal radius.
(We have adopted
a simple normalization for the AdS/black-hole metric, e.g., for 
$AdS^7$, ${\bar g}_{\tau\tau}=f(r)= r^2- r^{-4}$. This corresponds to
fixing the ``thermal-radius" $R_1=1/3$ so that $\beta=2\pi R_1
=2\pi/3$.) In general, $\beta$ serves as the mass scale in the strong coupling limit.

\section{Pomeron in Strong Coupling}


We have found~\cite{BMT2} that  there is indeed a rather remarkable
correspondence of the overall mass and spin structure between the
spectrum determined by lattice simulations at weak
coupling and that 
captured by the lowest states for fluctuations associated with the
type-IIA supergravity bosonic multiplet.  The glueball mass
calulations are based on the earlier work of several
authors~\cite{BMT1,others}. 
This supports the belief
  that the Maldacena conjecture may well be correct and that
further efforts to go beyond strong coupling are worthy of sustained
effort. We note that for each value of $PC = (++,-+,+-,--)$, the lowest state is
present in approximately the right mass range~\cite{BMT1,BMT2}.  

At higher masses, the discrepancies increase. One reason is the
obvious fact that on the supergravity side all orbital excitations of
higher spin states are pushed to infinity in strong coupling by virtue
of the divergent closed-string tension,
$\sigma =   \frac{16 \pi g^2 N}{ 27\beta^2} \; [\; 1
+ 0(\frac{1}{g^2 N})\; ] \;.
$
 For example, a $3^{++}$ state is a purely stringy effect outside of the
classical limit of supergravity. 

Finally, we must emphasize that our comparison is premised on the
neglect of many states in the strong coupling limit that are in the
wrong superselection sector to survive in the weak coupling limit of
QCD.  A major
challenge is to understand how this could be accomplished, e.g., 
by modifying or by adopting a differnt  background metric.

We next comment on the intercept of the leading
glueball trajectory as a way to estimate the crossover value for the
bare coupling, where continuum physics might begin to hold.
The Pomeron is the leading Regge trajectory passing through the
lightest glueball state with $J^{PC}=2^{++}$. In a linear
approximation, it can be parameterized by
$\alpha_P(t) = 2 + {\alpha'_P} (t-m_T^2), 
$
where  our strong-coupling result provides for the lightest tensor mass. 
The Pomeron slope can be related to the QCD string tension,
$\alpha_P'\simeq [ {27\over 32 \pi^2 g^2 N}+ 0(\frac{1}{g^4 N^2})] \; {
\beta^2}. 
$ 
 Putting these together,
we obtain a strong coupling expansion for the Pomeron intercept
\begin{equation}
\alpha_P(0) \simeq 2- 0.66 \; (\frac{4 \pi}{ g^2 N}) + 0(\frac{1}{g^4 N^2}) \; .
\end{equation}

Turning above  argument around would allow one to estimate a crossover value
between the strong and weak coupling regimes by fixing $\alpha_P(0)
\simeq 1.2$ at its phenomenological value~\cite{tan}.  In fact this
yields for $N = 3$ QCD a reasonable value for the fine structure
constant: $g^2/4 \pi \simeq 0.176$ at a characteristic confinement scale,
$\Lambda_{QCD}$.  Such estimates have proven sensible in the lattice
approach to strong coupling QCD.  Much more experience with this new
approach to strong coupling must be gained before such numerology can
be taken seriously. However if it proves to be an accurate guide, one
may be able to follow the general strategy used in lattice
simulations. Postpone the difficult question of analytically solving
the QCD string to find the true UV fixed point.  Work at a fixed but
physically reasonable cut-off scale and calculate the spectrum of QCD
in the cut-off theory. If one is near enough to the fixed point, mass
ratios should be reliable. After all, the real benefit of a weak/strong
duality is to use each method in the domain where it provides the
natural language.  On the other hand, clearly from a fundamental point
of view, finding analytical tools to understand the renormalized
trajectory and prove asymptotic scaling within the context of the
gauge invariant QCD string would also be a major achievement --- an
achievement that should include a proof of confinement
itself.

\section*{Acknowledgments}
I would like to thank 
Richard Brower and Samir Mathur for collaboration. This work was
supported by DOE grant DE-FG02/19ER40688-(Task-A).

\end{document}